\documentclass[a4paper,11pt]{article}
\usepackage{pos}

\begin{document}
\title{Neutrino nonradiative decay and the diffuse supernova neutrino background}
\author*[a]{Pilar Iváñez-Ballesteros}
\author[b]{Maria Cristina Volpe}
\affiliation[a]{Universite Paris Cite, Astroparticule et Cosmologie,\\ 10, rue A. Domon et L. Duquet, Paris, France}
\affiliation[b]{CNRS, Universite Paris Cite, Astroparticule et Cosmologie,\\ 10, rue A. Domon et L. Duquet, Paris, France}
\emailAdd{ivanez@apc.in2p3.fr}
\emailAdd{volpe@apc.univ-paris7.fr}
\abstract{The diffuse supernova neutrino background (DSNB) is the constant flux of neutrinos and antineutrinos emitted by all past core collapses in the observable Universe. We study the potential to extract information on the neutrino lifetime from the upcoming observation of the DSNB flux. The DSNB flux has a unique sensitivity to neutrino nonradiative decay for $\tau / m \in \left[ 10^9, 10^{11}\right]$~s/eV. To this end, we integrate, for the first time, astrophysical uncertainties, the contribution from failed supernovae, and a three-neutrino description of neutrino nonradiative decay. We present our predictions for future detection at the running Super-Kamiokande + Gd and the upcoming Hyper-Kamiokande, JUNO, and DUNE experiments. Finally, our results show the importance of identifying the neutrino mass ordering to restrict the possible decay scenarios for the impact of nonradiative neutrino decay on the DSNB.
}
\FullConference{XVIII International Conference on Topics in Astroparticle and Underground Physics (TAUP2023)\\
 28.08-01.09.2023\\
University of Vienna\\}
\maketitle

\section{Introduction}
When a massive star ($M \gtrsim 8 M_\odot$) dies, its inner core collapses and settles into either a neutron star (NS) or a black hole (BH). Most of the gravitational energy ($\sim 10^{53}$~erg) is released as neutrinos during this process. 
So far, the only observation of supernova neutrinos was made in 1987, when three experiments, Kamiokande-II~\cite{Kamiokande-II:1987idp}, IMB~\cite{Bionta:1987qt}, and Baksan~\cite{Alekseev:1988gp} observed 24 neutrino events associated with the explosion of SN1987A. 
However, core collapses in our galaxy or its proximities are very rare, with a rate of only a few events per century~\cite{Rozwadowska:2020nab}. 

The diffuse supernova neutrino background (DSNB) provides an alternative for the observation of supernova neutrinos. The DSNB is the flux of all neutrinos and antineutrinos emitted by all past core collapses in the observable Universe. Although the DSNB flux has not been detected yet, several experiments have set upper limits to the $\bar\nu_e$~\cite{Super-Kamiokande:2021jaq}, the $\nu_e$~\cite{SNO:2020gqd}, and the non-electronic, $\nu_x$,~\cite{Lunardini:2008xd} fluxes. The next generation of neutrino experiments is expected to observe the DSNB flux. Experiments like SK loaded with gadolinium (SK-Gd), which is already running, and the forthcoming Hyper-Kamiokande (HK) and JUNO experiments will be sensitive to the DSNB $\bar\nu_e$ flux through inverse beta decay. On the other hand, the Deep Underground Neutrino Experiment (DUNE) will be able to detect the $\nu_e$ flux using neutrino absorption in liquid argon.

The detection of the DSNB flux will provide complementary information to the detection of the neutrino flux from a single supernova since the DSNB contains information about the star-formation rate and the fraction of failed supernovae~\cite{Lunardini:2009ya}. The DSNB flux will also provide information on nonstandard properties like neutrino decay. The DSNB flux has a unique sensitivity to nonradiative neutrino decay for $\tau/m \in [10^9, 10^{11}]$~s/eV. Here we summarize the main findings of our work \cite{Ivanez-Ballesteros:2022szu}. We investigated the effect of neutrino decay on the DSNB flux considering a three-flavor neutrino framework and a detailed astrophysical model in which we included the uncertainties from the evolving core-collapse rate and the contributions from failed supernovae.

\section{Description of the diffuse supernova neutrino background}

In the absence of neutrino decay, the DSNB flux of the mass eigenstate $\nu_i$ has this expression:
\begin{equation}
\label{eq:dsnbflux}
\phi_{\nu_i}(E_{\nu}) = {c} \int_0^{z_\mathrm{max}} \frac{dz}{H(z)} \int_{8~ \rm M_\odot}^{125~\rm M_\odot} {d \rm M}~R_{\rm SN}(z,{\rm ~M}) ~ {Y_{\nu_i}(E'_{\nu}, {\rm ~M})} \ ,
\end{equation}
where $E'_\nu = E_\nu (1+z)$ is the redshifted neutrino energy; $c$, the speed of light; and $z$, the cosmological redshift. We take $z_{\rm max} = 5$. In the expression above, $H(z)$ represents the Hubble rate which is defined by $H(z) = H_0 \left(\Omega_m (1+z)^3 + \Omega_\Lambda\right)^{1/2}$. In our calculations, we take the Hubble constant $H_0 = 67.4$~km~s$^{-1}$ and the matter and dark energy densities as $\Omega_m = 0.3$ and $\Omega_\Lambda = 0.7$, respectively.

In Eq.~(\ref{eq:dsnbflux}), the term $R_{\rm SN}(z,{\rm ~M})$ indicates the evolving core-collapse supernova rate which is proportional to the star-formation rate. In our work, we take the star-formation rate described in Ref.~\cite{Yuksel:2008cu}. It is important to note that the local core-collapse supernova rate has a large uncertainty that impacts the DSNB and constitutes one of its largest uncertainties.

The last important input in Eq.~(\ref{eq:dsnbflux}) is the neutrino flux emitted by a single core collapse, $Y_{\nu_i} (E'_\nu, \rm M)$. Neutrinos are produced in the inner core and start free streaming at the neutrinosphere. The time-integrated flux at the neutrinosphere can be described using a power-law distribution~\cite{Keil:2002in}.
%
%
We use the results obtained by one-dimensional simulations by the Garching group~\cite{Priya:2017bmm}.

After being emitted at the neutrinosphere, neutrinos travel through matter until they reach the surface of the supernova, experiencing the Mikheev-Smirnov-Wolfenstein (MSW) effect during this propagation~\cite{Wolfenstein:1977ue, Mikheev:1986wj}.
Flavor conversion inside the supernova is still under study, and it can also be affected by $\nu$-$\nu$ interactions, shock wave effects, or turbulence (see Ref.~\cite{Volpe:2023met} for a comprehensive review). However, for simplicity, we only consider here the well-established MSW effect.


The spectrum of the emitted neutrino flux depends on different aspects, such as the progenitor's mass or the outcome of the collapse. In particular, BH-forming collapses emit a hotter spectrum than NS-forming collapses. Therefore, although subdominant, the contribution from ``failed" supernovae can significantly affect the DSNB flux~\cite{Lunardini:2009ya} and, especially, the tail of the flux. 

\section{Neutrino nonradiative decay}
Let us now investigate the effect of nonradiative neutrino decay on the DSNB flux.
We consider the process: $\nu_i \rightarrow \nu_j (\bar\nu_j) + \phi$,
in which a heavier neutrino $\nu_i$ decays into a lighter (anti)neutrino $\nu_j (\bar\nu_j)$ and a light or massless scalar particle, $\phi$~\cite{Ando:2003ie}. 

\begin{figure}[t]
\begin{center}
\includegraphics[scale=0.4]{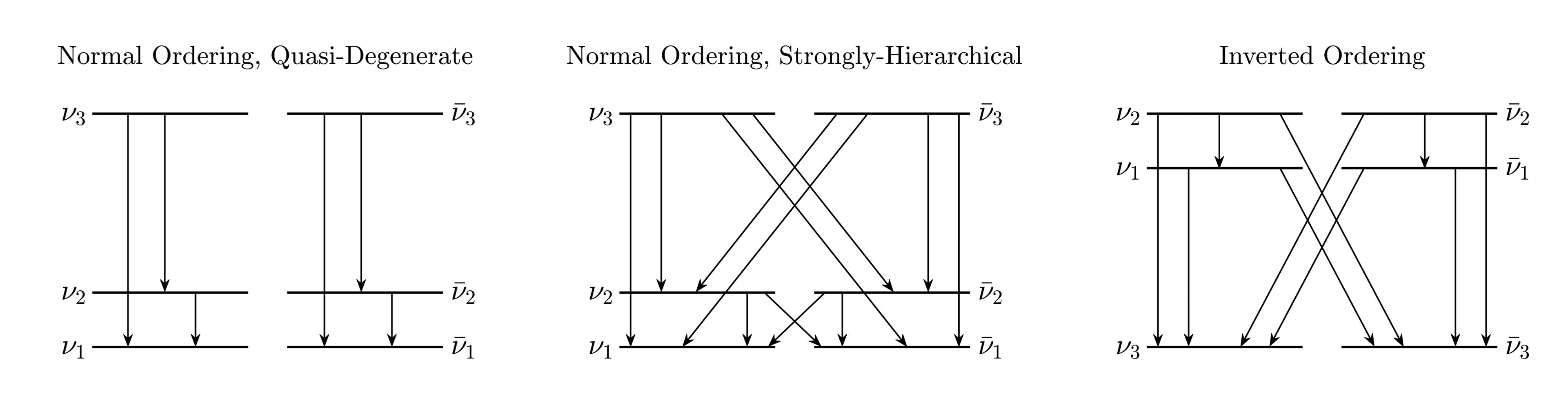}
\caption{Decay patterns (Figure taken from Ref.~\cite{Ivanez-Ballesteros:2022szu}). Left:  Normal ordering in the quasi-degenerate case (QD).
Middle: Normal ordering in the strongly hierarchical (SH) case. Right:  Inverted ordering case. 
}
\label{fig:decaypatterns}
\end{center}
\end{figure}

\begin{figure}[!t] 
    \centering
        \includegraphics[scale=0.32]{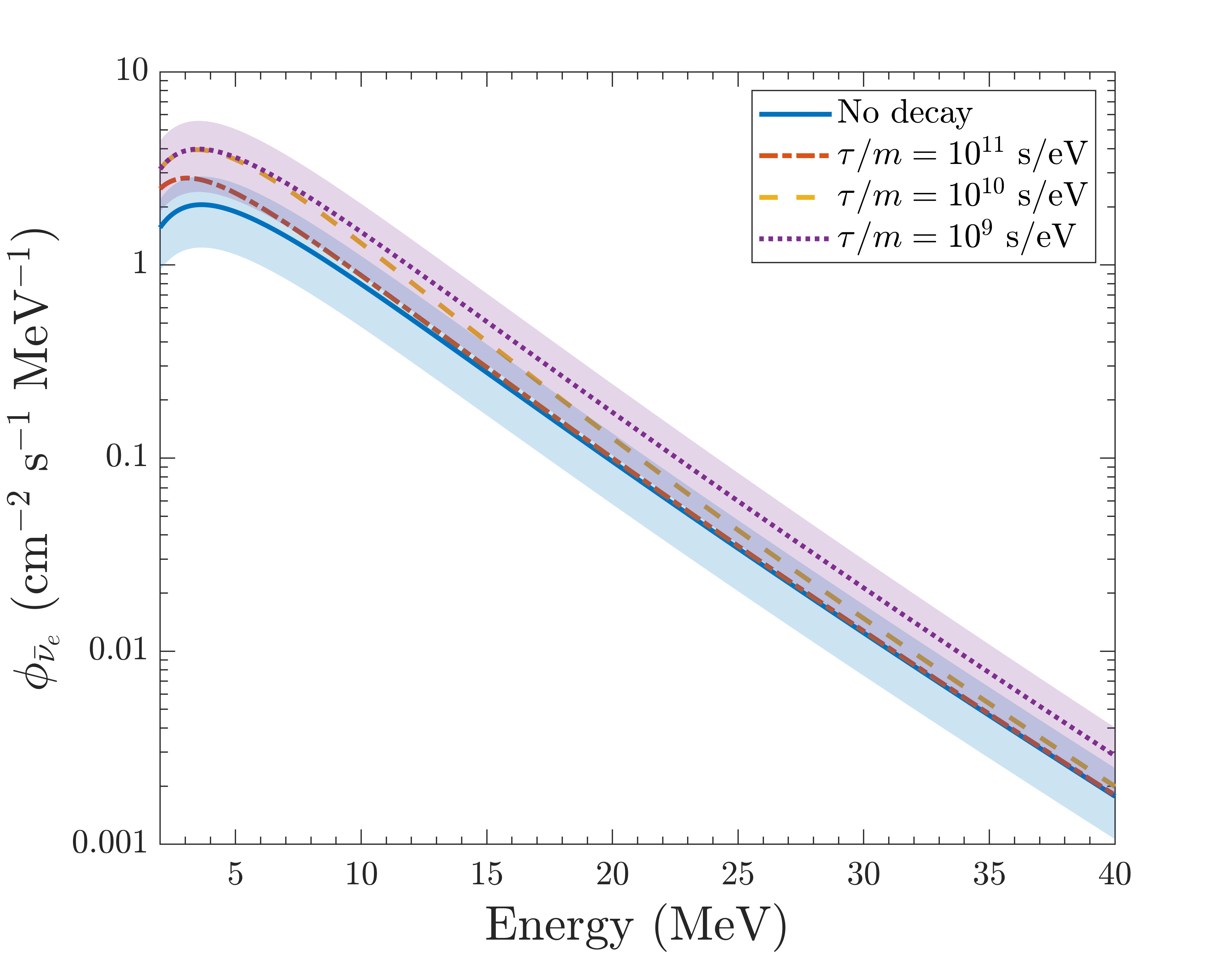}    
        \includegraphics[scale=0.32]{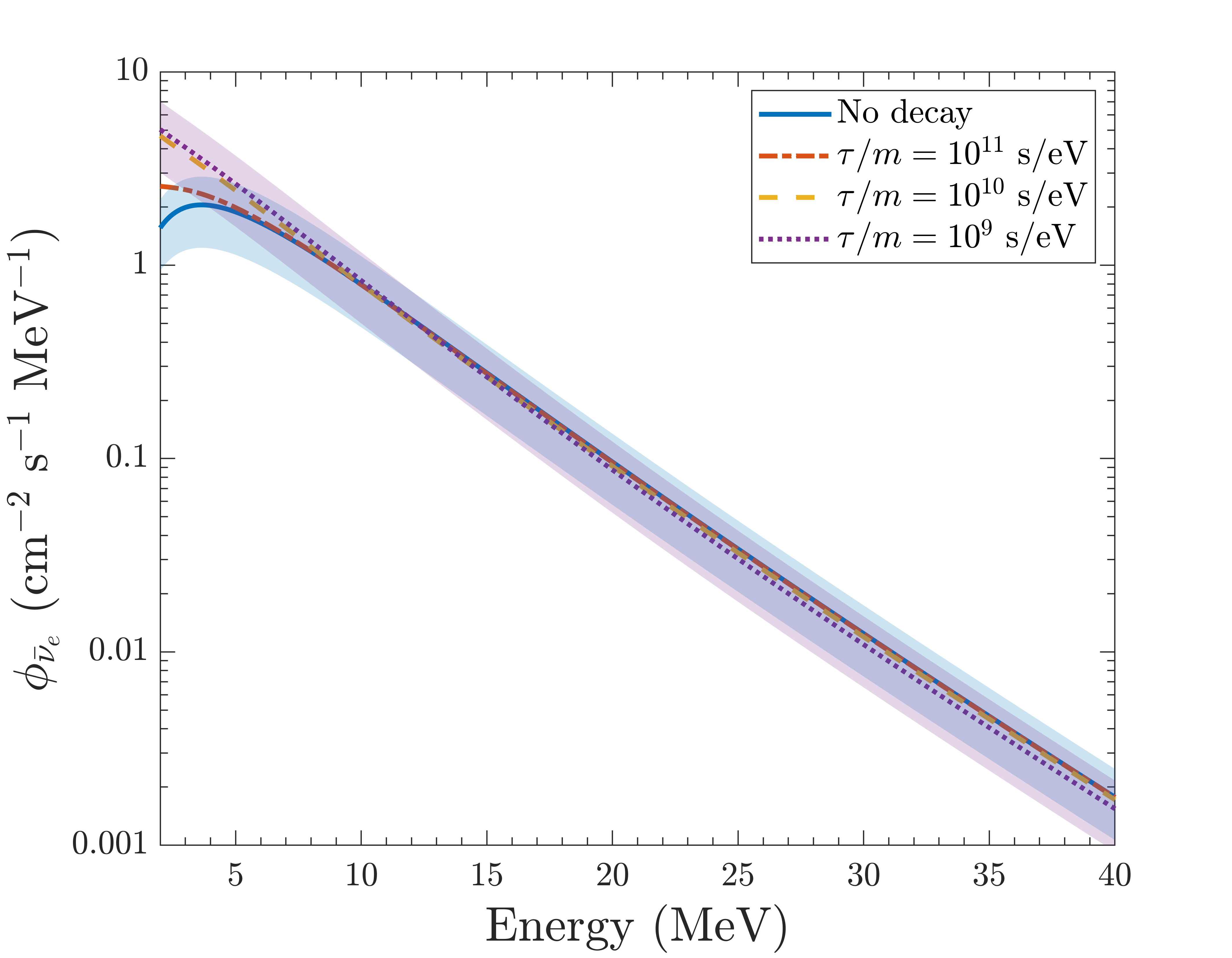}
        \caption{Results for the DSNB $\bar\nu_e$ flux in the presence of neutrino nonradiative decay for the NO, QD case (left) and the NO, SH case (right) (Figures taken from Ref.~\cite{Ivanez-Ballesteros:2022szu}). The results are obtained considering our {\it fiducial} model for different values of $\tau /m$ and the case of no decay. In the cases of no decay and complete decay ($\tau /m = 10^9$~s/eV), the bands indicate the uncertainty from the core-collapse supernova rate.} 
 \label{fig:fluxes_NO}
\end{figure}

Neutrino decay has already been studied and several limits have been set. Ref.~\cite{Ivanez-Ballesteros:2022szu} summarizes the main results from terrestrial experiments, astrophysical sources, and cosmological observables. 
In our recent study~\cite{Ivanez-Ballesteros:2023lqa}, we obtained in the case of IO the lower bound $\tau/m \geq 1.2 \times 10^5$~s/eV at 90\% CL for the lifetime-to-mass ratio of $\nu_1$ and $\nu_2$ by performing a 7-dimensional likelihood analysis of the 24 SN1987A events detected in Kamiokande-II, IMB, and Baksan. The effects of nonradiative neutrino decay on the DSNB flux were investigated in previous studies~\cite{Ando:2003ie,Fogli:2004gy,DeGouvea:2020ang, Tabrizi:2020vmo}. However, previous works lacked a three-flavor neutrino framework or a detailed astrophysical model.

In the present study, we consider both NO and IO since the mass ordering of neutrinos remains unknown. 
Fig.~\ref{fig:decaypatterns} represents the three decay patterns we consider: the quasi-degenerate (QD) case ($m_3\simeq m_2 \simeq m_1$) and the strongly hierarchical (SH) case ($m_3 \gg m_2 \gg m_1 \simeq 0$) in NO, and the IO case ($m_2 \simeq m_1 \gg m_3$). We use a {\it democratic} hypothesis for the branching ratios, $B(\nu_i \rightarrow \nu_j)$ (see Fig.~3 of Ref.~\cite{Ivanez-Ballesteros:2022szu}) and we assume equal values of $\tau/m$ for the decaying mass eigenstates.

\section{Predictions for the DSNB fluxes and events with/without neutrino decay}
\begin{figure}[!t] 
    \centering
        \includegraphics[scale=0.32]{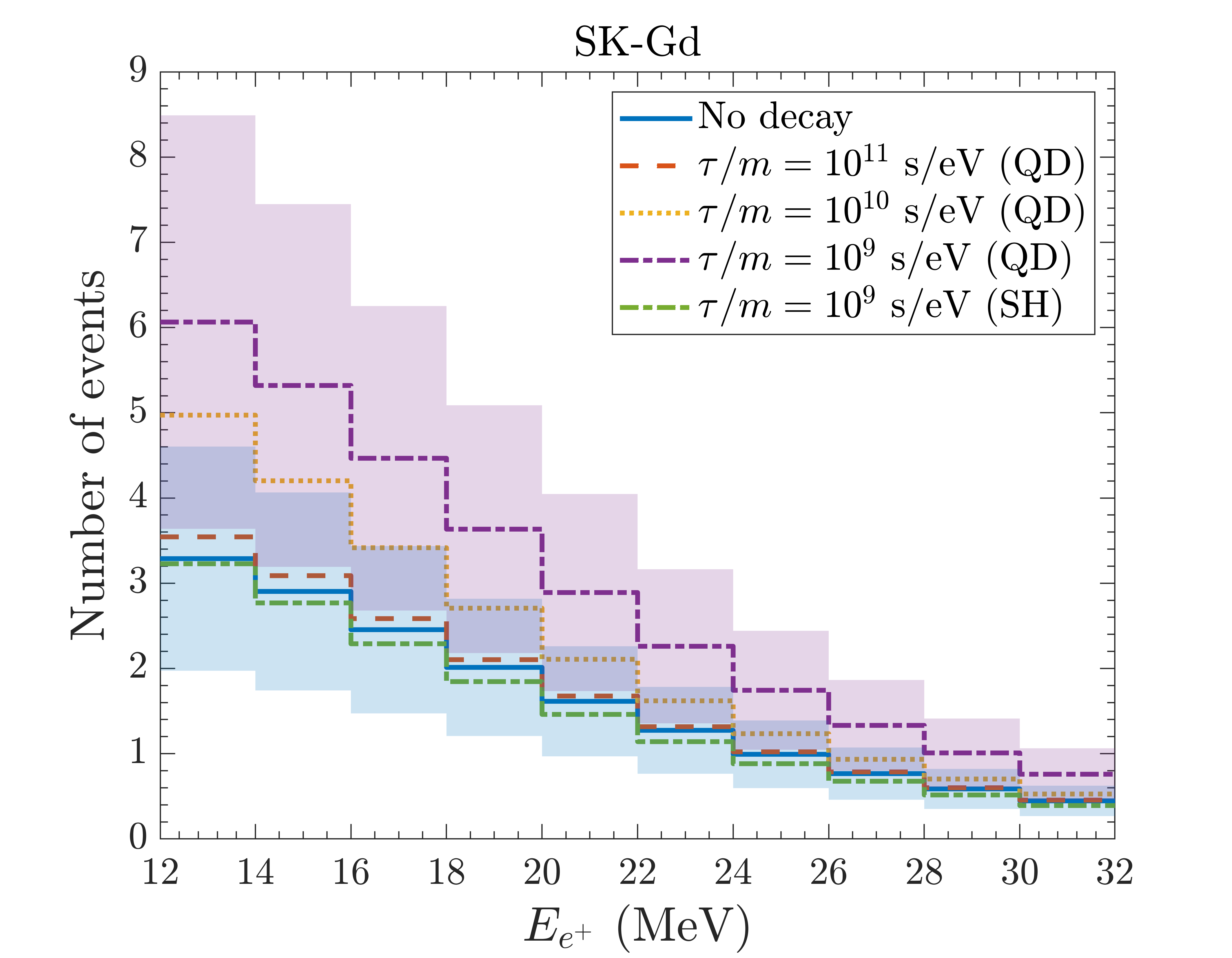}    
        \includegraphics[scale=0.32]{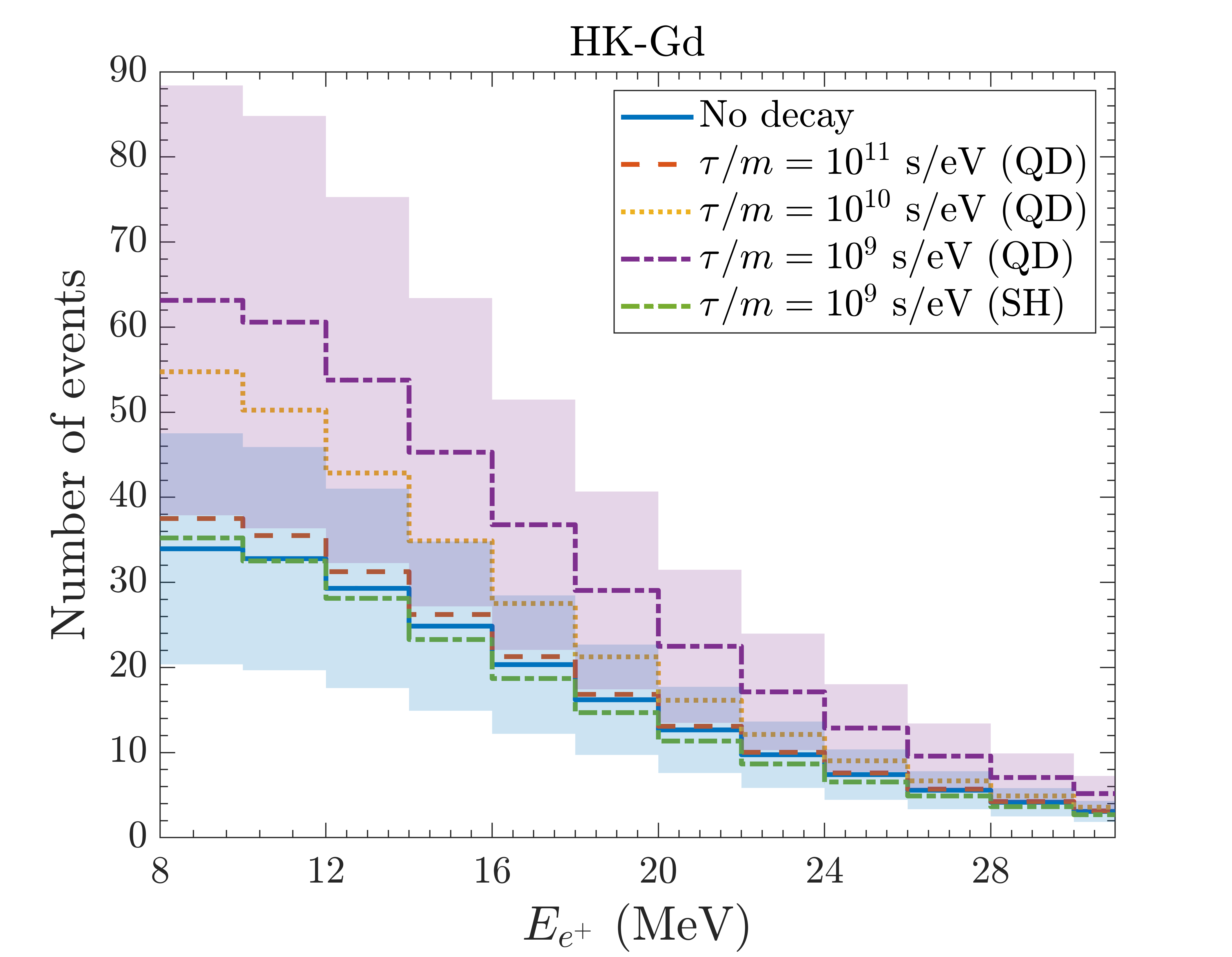}
        \includegraphics[scale=0.32]{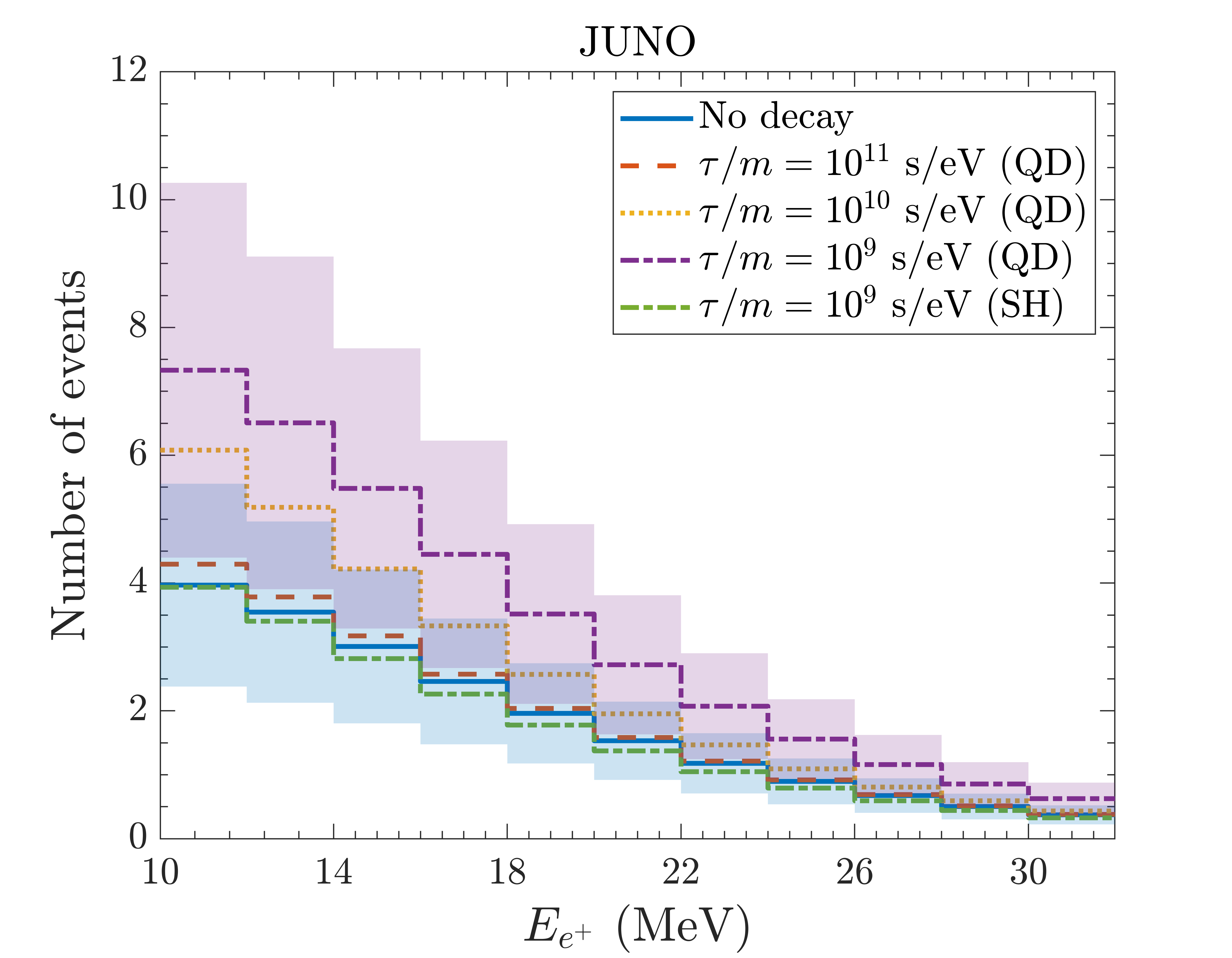}    
        \includegraphics[scale=0.32]{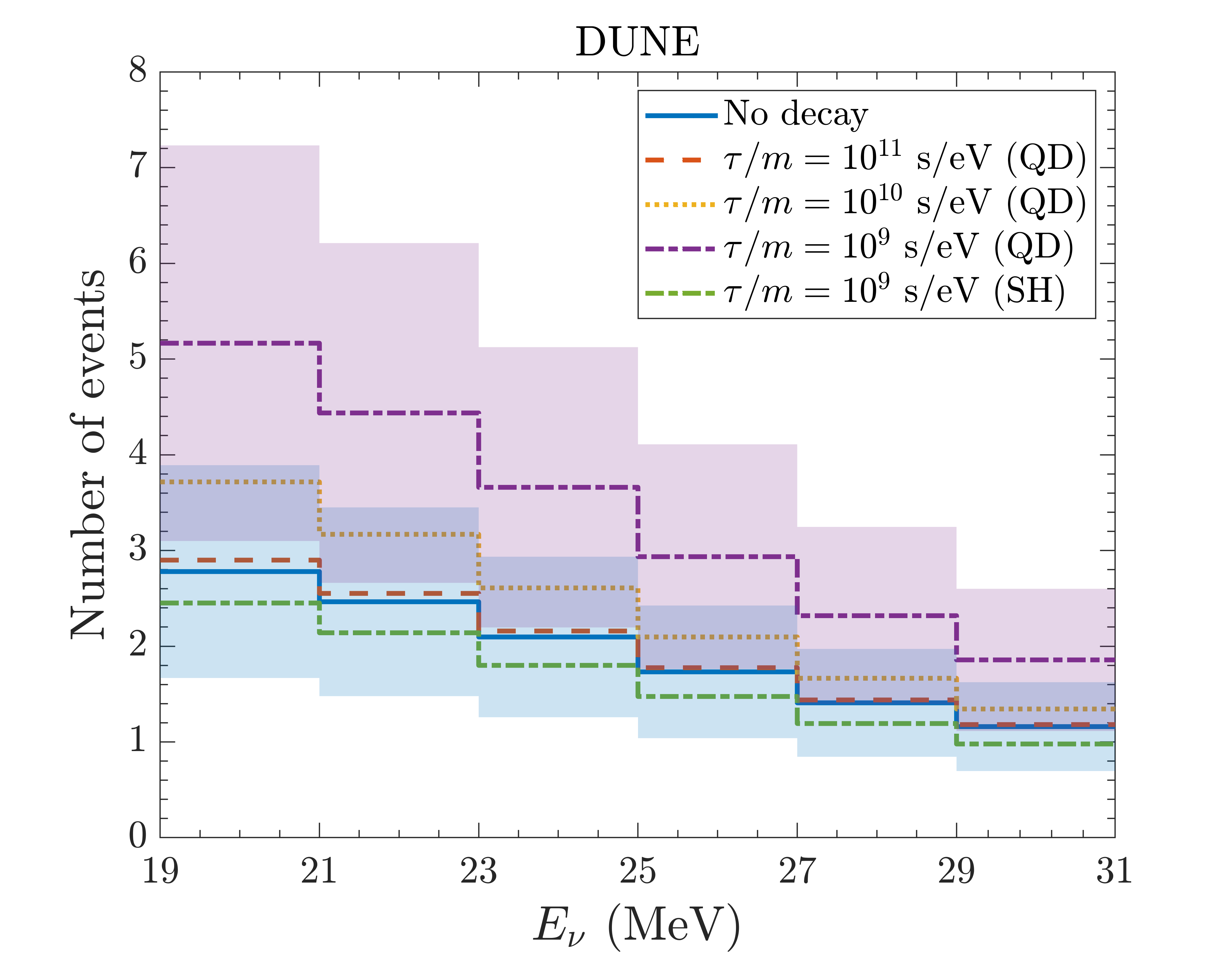}   
        \caption{Expected DSNB events considering NO at SK-Gd (top left), HK-Gd (top right), JUNO (bottom left), and DUNE (bottom right) after 10 years of exposure for SK-Gd and 20 years of exposure for the rest of the experiments (Figures taken from Ref.~\cite{Ivanez-Ballesteros:2022szu}). The results are obtained considering our {\it fiducial} model for different values of $\tau /m$ and the case of no decay. In the cases of no decay and complete decay ($\tau /m = 10^9$~s/eV), the bands indicate the uncertainty from the core-collapse supernova rate.} 
 \label{fig:Events_NO}
\end{figure}

\begin{figure}[!htb] 
    \centering
        \includegraphics[scale=0.32]{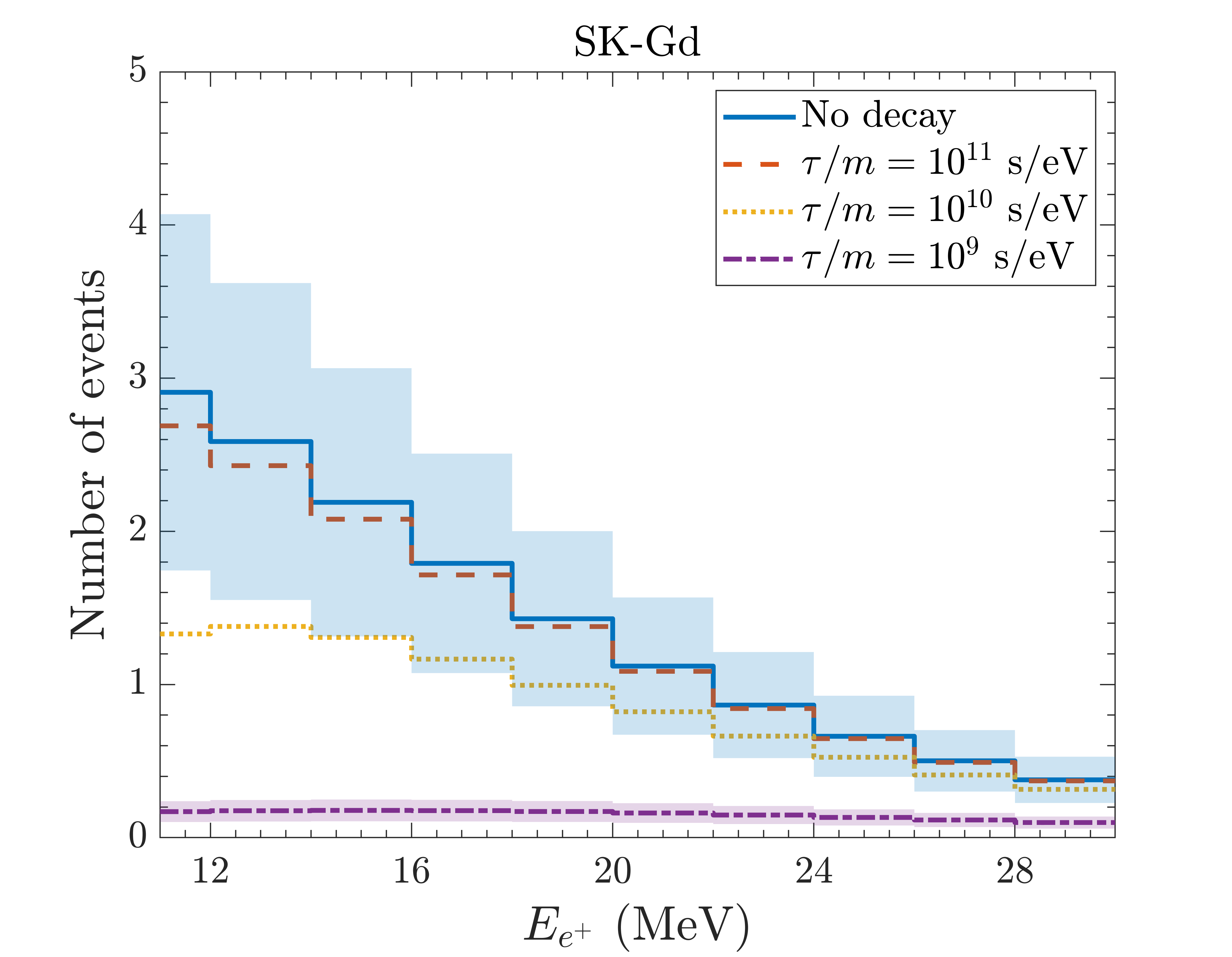}    
        \includegraphics[scale=0.32]{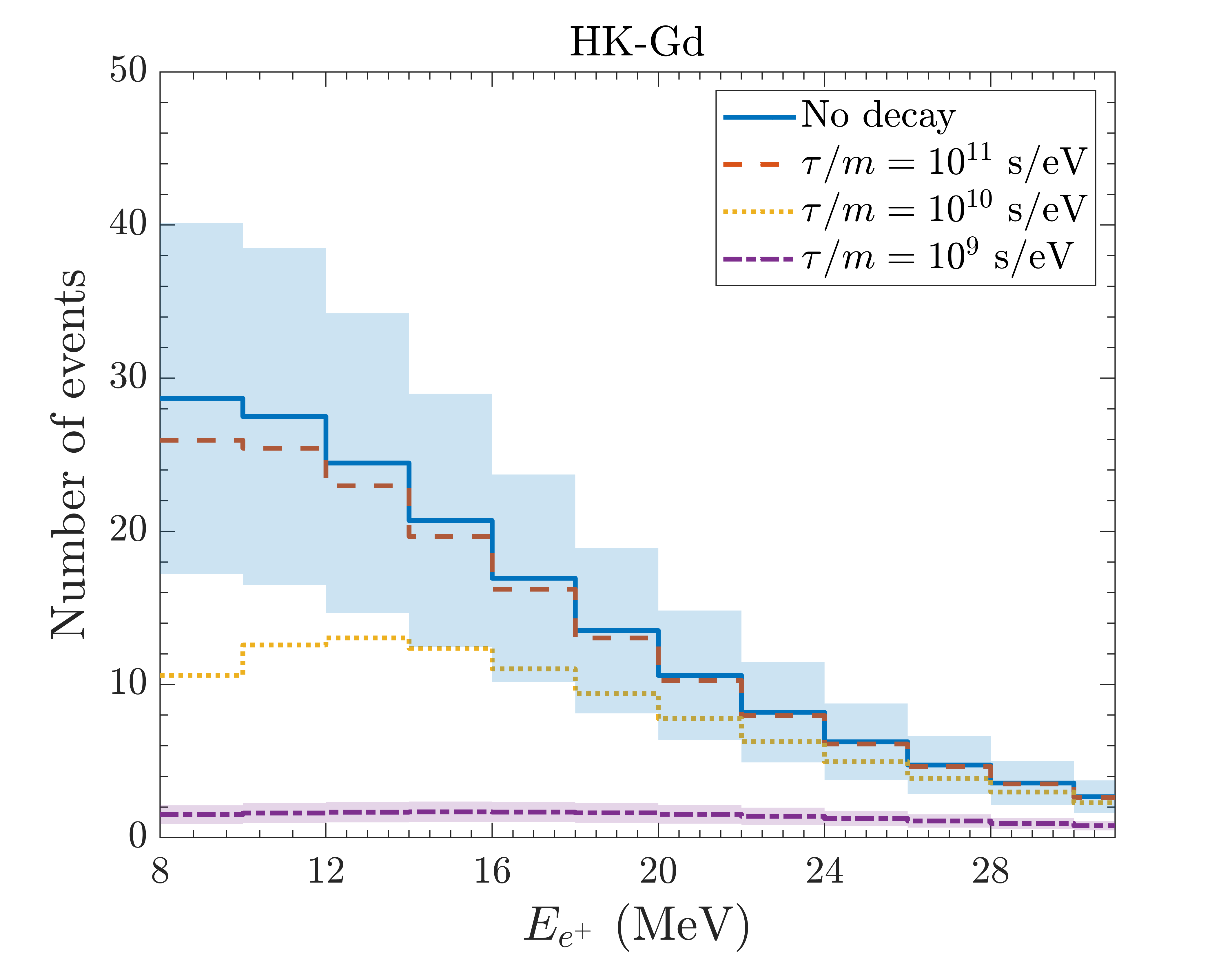}
        \includegraphics[scale=0.32]{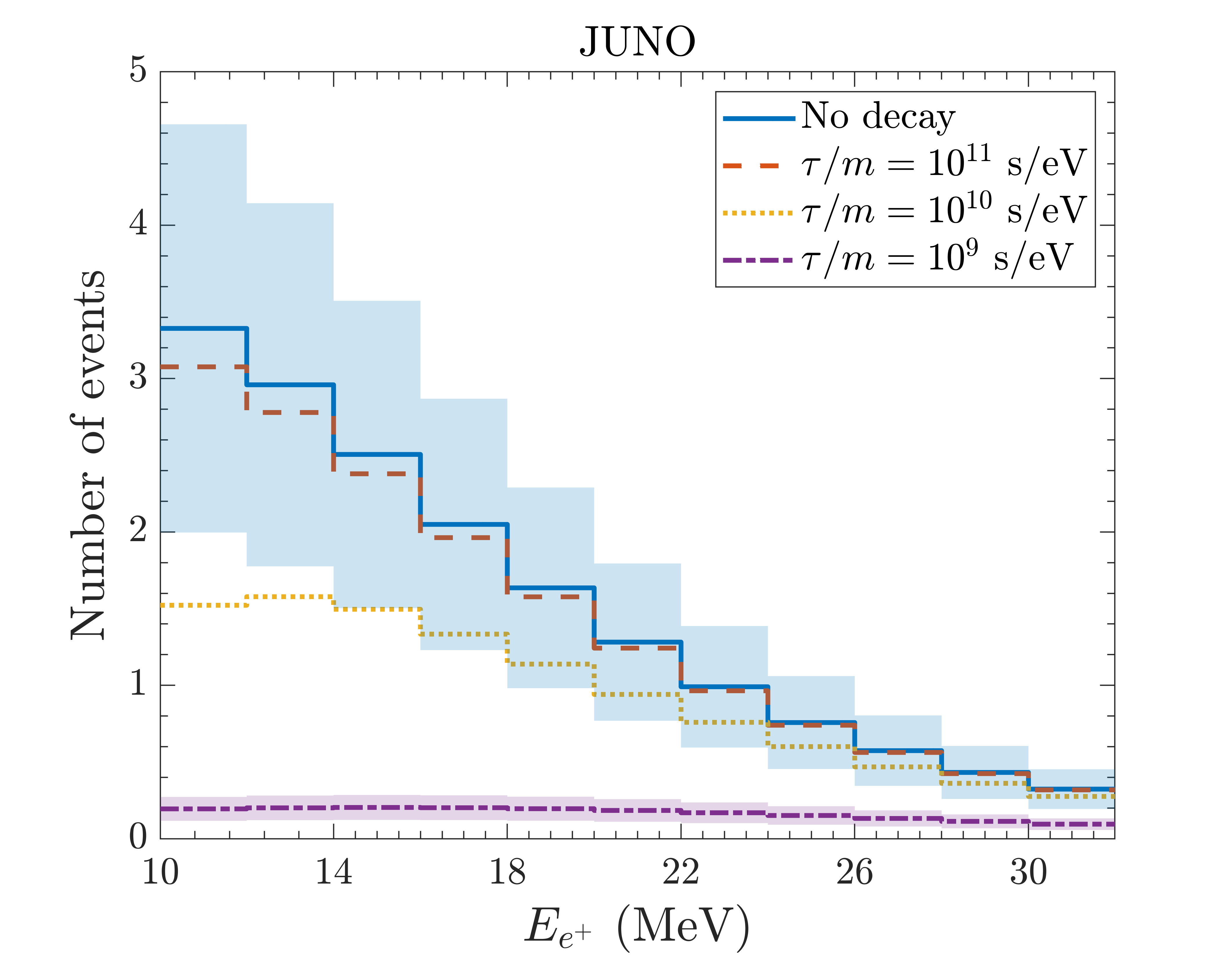}    
        \includegraphics[scale=0.32]{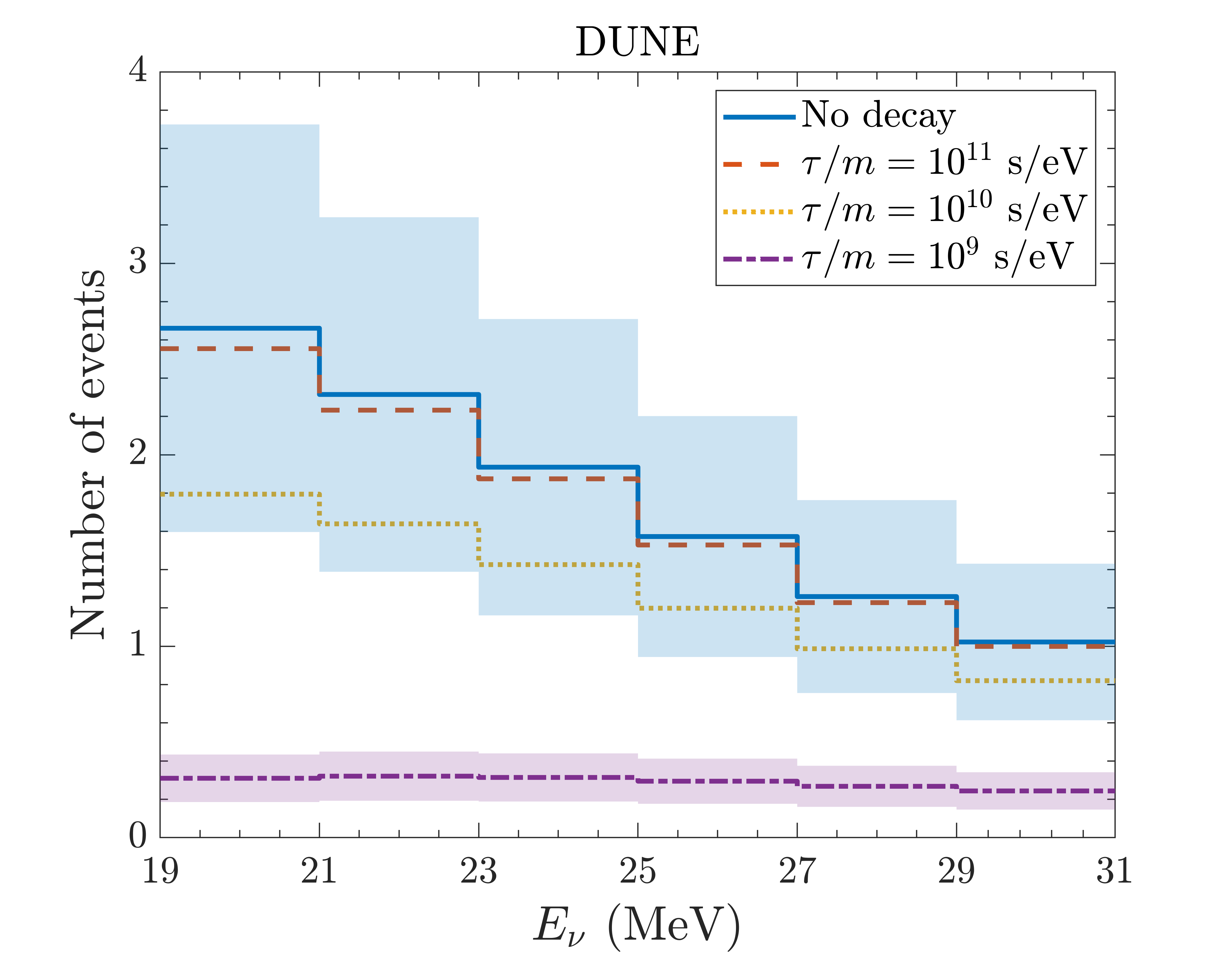}
        \caption{Same as Fig.~\ref{fig:Events_NO} but in the case of IO.}
 \label{fig:Events_IO}
\end{figure}

First, let us comment on the results for the DSNB flux in the absence of neutrino decay. Since the fraction of BH-forming collapses, $f_{\rm BH}$, is still unknown, following Refs.~\cite{Priya:2017bmm, Moller:2018kpn}, we consider three scenarios: $f_{\rm BH} =$ 0.09, 0.21, 0.41 \cite{Ivanez-Ballesteros:2022szu}. The last two cases are consistent with detailed supernova simulations \cite{Kresse:2020nto}. Our {\it fiducial} model assumes $f_{\rm BH} = 0.21$ and the core-collapse rate of Ref.~\cite{Yuksel:2008cu}. The DSNB fluxes for $\bar\nu_e$ and $\nu_e$ on Earth for these scenarios are shown in Fig.~4 of Ref.~\cite{Ivanez-Ballesteros:2022szu}. For the integrated fluxes, for our {\it fiducial} model
, we obtain in NO (IO): 
\begin{align*}
    \phi_{\bar\nu_e}& (E_\nu > 17.3~{\rm MeV}) = 0.77 \pm 0.30~(0.63 \pm 0.25)~{\rm cm}^{-2}\rm s ^{-1}, \\
    \phi_{\nu_e}& (22.9 < E_\nu < 36.9~{\rm MeV}) = 0.20 \pm 0.08~(0.18 \pm 0.08)~{\rm cm}^{-2}\rm s ^{-1}.
\end{align*}
The results for the $\bar\nu_e$ flux are below the current upper limit
obtained from the combined analysis of SK data~\cite{Super-Kamiokande:2021jaq} by a factor of $\sim 4$. On the other hand, our results for the integrated $\nu_e$ flux are two orders of magnitude lower than the upper limit set by SNO data 
\cite{SNO:2020gqd}.

Using the expressions shown in the Appendix of Ref.~\cite{Ivanez-Ballesteros:2022szu}, we obtain the DSNB fluxes in the presence of neutrino decay. Fig.~\ref{fig:fluxes_NO} shows the results for the $\bar\nu_e$ flux in the case of NO: for QD masses (left) and SH masses (right). In the QD case, the presence of decay enhances the $\bar\nu_e$ flux. This enhancement is greater the shorter the lifetime is. On the other hand, in the SH case, the fluxes with and without decay are almost degenerate at higher energies and differ only at low energies, below current thresholds. The results for the flux in the case of IO can be found in Ref.~\cite{Ivanez-Ballesteros:2022szu}. In this case, the fluxes present a strong suppression for short lifetimes with respect to the no decay case.

Our predictions for the number of events at SK-Gd, HK 
doped with gadolinium (HK-Gd), JUNO, and DUNE are shown in Fig.~\ref{fig:Events_NO} for NO and Fig.~\ref{fig:Events_IO} for IO. 
For the NO, QD case (Fig.~\ref{fig:Events_NO}), the number of events is larger for shorter lifetimes. On the other hand, for SH masses, we observe that with $\tau/m = 10^9$~s/eV, the results are almost degenerate with respect to the case of no decay. The astrophysical uncertainties make the two cases indistinguishable. 
In the IO case (Fig.~\ref{fig:Events_IO}), the number of events is suppressed when considering neutrino decay. 
In particular, for $\tau/m = 10^{9}$~s/eV, the events are highly suppressed and are clearly distinguishable from the no decay case. 


Our results highlight the important implications that the determination of the mass ordering will have on the DSNB predictions in relation to nonradiative decay.
Indeed, if the mass ordering is inverted, the non-observation of the DSNB could be due to nonradiative neutrino decay instead of conservative inputs from standard physics.
If the mass ordering is normal, there are significant degeneracies in the DSNB predictions in the presence or absence of decay. Reducing
astrophysical uncertainties remains crucial to extracting the most information from the upcoming DSNB discovery.

\end{document}